\documentclass[aps,pre,showpacs,raggedbottom, 
amssymb,twocolumn,groupedaddress]{revtex4-1}
\usepackage{graphicx}
\usepackage{amsmath}
\usepackage{amsfonts}
\usepackage{amssymb}
\usepackage{epsfig,libertine}
\usepackage[libertine]{newtxmath}
\usepackage{color}
\usepackage{dsfont, hyperref, xcolor}
\usepackage{comment}
\usepackage{enumerate}

\newcommand{\abs}[1]{\left\vert#1\right\vert}

\newcommand{\bra}[1]{\langle#1\vert}
\newcommand{\ket}[1]{\vert#1\rangle}

\newcommand{\sign}{\textrm{sign}}
\newcommand{\change}{\color{black}}

\begin{document}

\title{\change Correlations, long-range entanglement and dynamics 
in long-range Kitaev chains}

\author{Gianluca~Francica, Luca Dell'Anna}
\address{Dipartimento di Fisica e Astronomia 
e Sezione INFN, 
Universit\`{a} 
di Padova, via Marzolo 8, 35131 Padova, Italy}

\date{\today}

\begin{abstract}
Long-range interactions exhibit surprising features which have been less explored so far. Here, studying a one-dimensional fermionic chain with long-range hopping and pairing, we discuss some general features associated to the presence of long-range entanglement. In particular, after determining  the algebraic decays of the correlation functions, we prove that a long-range quantum mutual information exists if the exponent of the decay is not larger than one. Moreover, we show that the time evolution {\change triggered by a quantum} 
quench between short-range and long-range regions, can be characterized by  dynamical quantum phase transitions {\change without crossing any phase boundary. We show, also, that 
the adiabatic dynamics is dictated by the divergence of a topological length scale at the quantum critical point, clarifying the violation of the Kibble-Zurek mechanism for long-range systems.}
\end{abstract}

\maketitle

\section{Introduction}
The study of the correlations between the parties of a many-body system in its quantum phases is a fundamental problem of condensed matter physics. A special role is played by the correlations that cannot be generated by local unitary evolutions, which give a so-called long-range entanglement~\cite{chen10}. In two dimensional systems, this leads to extraordinary topological phenomena, such as topological order with topological degeneracy of the quantum phase~\cite{wen90} and anyon excitations which can be employed in quantum fault-tolerant computation purposes~\cite{kitaev03}, whose origin is revealed by topological entanglement entropy~\cite{kitaev06,levin06}.
On the contrary, it is well known that the gapped quantum phases of one-dimensional systems can show only short-range entanglement if the interactions are short-ranged.
Typically, for this kind of interactions, short-range and long-range entanglement regions are separated by a energy gap closing.
Things change drastically when we consider long-range interactions. In this case, there is a path in the parameter space connecting short-range and long-range entanglement regions without closing the gap, also in one-dimensional systems. 
Moreover, the conventional topological classification~\cite{Schnyder08,kitaev09} cannot be applied in the presence of long-range interactions, and new topological features emerge, like the presence of massive Dirac edge modes~\cite{Viyuela06,lepori17}.

Here, we prove that the long-range entanglement in one-dimensional systems is intimately related to the algebraic decay of the correlation functions. In doing this, we characterize the long-range entanglement in the bulk by using the mutual information between two subsystems with sizes and separation which linearly increase with the size of the system.
This quantity counts the total amount of correlations \cite{groisman} and is an upper bound for squared correlation functions \cite{wolf}. 
We find that it does not vanish in the thermodynamic limit if the correlation functions decay algebraically with an exponent not larger than one. 
We {\change discuss} our results with the help of a specific model that is a chain of fermions with long-range hopping and pairing~\cite{vodola14,alecce17,lepori17,jager20}, which we {\change investigate} thoroughly. 
{\change Experimental realizations of long-range topological superconductors employ one-dimensional arrays of magnetic impurities on top of a conventional superconducting substrate \cite{perge,pawlak,ruby}, leading to the realization of an effective Kitaev Hamiltonian with both long-range pairing and hopping.} 
The model displays {\change some peculiarities} including continuous quantum phase transitions without mass gap closure, violation of the area-law for the von Neumann entropy and emergence of massive edge states.
For this model, we {\change calculate} analytically the asymptotic formulas for the {\change algebraic} decay of the correlation functions {\change for any value of the long-range couplings.} 
To make sure of the results, we also {\change characterize} the long-range entanglement numerically by analyzing the largest Schmidt eigenvalue {\change finding the effective central charge also in the presence of long-range hopping. 
Taking advantage of the correlation functions we, then, find the long-distance behavior of the mutual information shared by two disjoint segments of the chain.} 
Finally, by considering the time evolution generated by a quantum quench between short-range and long-range entanglement regions, we {\change prove} that there are dynamical quantum phase transitions~\cite{heyl13,heyl18}, 
{\change while, in the adiabatic regime, we show how the Kibble-Zurek mechanism~\cite{kibble76,zurek85} is related to a topological scale length~\cite{cheng17} at the quantum critical point.}

{\change The paper is organized as follows: we begin introducing the long-range Kitaev chain which can host algebraically localized Majorana modes at the edges, reporting in Appendix a general improved method, introduced in Ref. ~\cite{jager20}, useful to detect the Majorana zero modes also in the presence of long-range couplings, and discussing its limit of validity. 
In Sec.~III, we provide a complete analysis of the correlation functions of the model, by introducing a different approach with respect to that used in Ref.~\cite{vodola14}, showing clearly that the origin of exponential and algebraic decays are related to poles and brunch cut of a complex integrand. In this way we generalize the previous results \cite{vodola14}, performing the analytical calculation for all possible sets of long-range couplings. In Sec.~IV, we discuss the long-range entanglement in one-dimensional systems from a general point of view, proving how its presence is related to the decay of correlation functions. Surprisingly we show that the mutual information shared by two disjoint regions can survive at infinite distances. In Sec.~V we complete the characterization of the long-range regime looking at some dynamical properties driven by sudden and adiabatic quantum quenches, proving the existence of dynamical phase transitions while explaining the peculiar adiabatic dynamics \cite{defenu19} in terms of a topological characteristic length scale \cite{cheng17}. 
We summarize our results in the final Section.}

\section{The model}

We consider 
the following fermionic Hamiltonian
\begin{eqnarray}
\nonumber H&=&-\frac{w}{2}\sum_{j=1}^L \sum_{l=1}^{L-1} u_l (a_j^\dagger a_{j+l}+h.c.)-\mu \sum_{j=1}^L(n_j-\frac{1}{2})\\
&& +\frac{\Delta}{2}\sum_{j=1}^L \sum_{l=1}^{L-1} v_l (a_j a_{j+l}+h.c.)\,,
\end{eqnarray}
where $a_j$ ($a^\dagger_j$) annihilates (creates) a fermion in the site $j$, $w$ is the hopping amplitude, $\mu$ is the chemical potential, $\Delta$ is the superconductive pairing. We consider an algebraic decay of hopping and pairing {\change couplings}, so that for a closed chain (with periodic boundary conditions), we consider $u_l=\theta(L/2-l)l^{-\alpha}+\theta(l-L/2)(L-l)^{-\alpha}$ and  $v_l=\theta(L/2-l)l^{-\beta}-\theta(l-L/2)(L-l)^{-\beta}$, 
{\change where $\theta$ is the Heaviside step function.} 
For an open chain, the sum with respect to $l$ runs from $1$ to $L-j$, and we consider $u_l = 2 l^{-\alpha}$ and $v_l=2 l^{-\beta}$. In particular, in the limit $\alpha\to \infty$ and $\beta\to \infty$ we recover the conventional short-range Kitaev chain~\cite{kitaev01}.

For a closed chain, we can perform a Fourier transform $a_j = \frac{1}{\sqrt{L}} \sum_k e^{-i k j} a_k$, with $k= 2\pi n/L$,  $n = -(L-1)/2,\cdots, (L-1)/2$ for $L$ odd, and  $n = -L/2+1,\cdots, L/2$ for $L$ even. By defining the Nambu spinor $\Psi_k = (a_k,a_{-k}^\dagger)^T$, the Hamiltonian reads
\begin{equation}\label{eq. hami k}
H = \frac{1}{2}\sum_k  \Psi_k^\dagger [ -(\mu+w\, g(k) )\tau_3 +\Delta\, f(k) \tau_2] \Psi_k\,,
\end{equation}
where $\tau_i$ with $i=1,2,3$ are the Pauli matrices and where we defined the following functions $g(k)=\sum_{l=1}^{L-1} u_l \cos( kl)$ and $f(k) = \sum_{l=1}^{L-1} v_l \sin( kl)$.  
The Hamiltonian in Eq.~\eqref{eq. hami k} can be written as $H=\sum_k \Psi^\dagger_k \vec d_k \cdot \vec \tau \,\Psi_k$, which, in the diagonal form, reads $H=\sum_k \epsilon_k \alpha_k^\dagger \alpha_k $, obtained after performing a rotation with respect the $x$-axis with an angle $\theta_k$ between $\vec d_k$ and the $z$-axis, corresponding to the Bogoliubov transformation $\alpha_k = \cos (\theta_k/2) a_k -i \sin(\theta_k/2) a^\dagger_{-k}$, where $\epsilon_k =2 ||\vec d_k||$, or more explicitly, 
{\change 
\begin{equation}
\change \epsilon_k =\sqrt{(\mu+w\, g(k))^2 +(\Delta\, f(k))^2}.
\end{equation}
}
In the thermodynamic limit, the functions $g(k)$ and $f(k)$ can be written in terms of polylogarithms  as  
{\change 
\begin{eqnarray}
g(k) = 2\, \textrm{Re} [Li_\alpha(e^{ik})],\\
f(k) = 2\, \textrm{Im} [Li_\beta(e^{ik})].
\end{eqnarray} 
As already mentioned, this model can host Majorana modes exponentially or algebraically localized at the edges, depending on the coupling parameters. An approach introduced in Ref.~\cite{jager20}, useful to find the spatial profile of the Majorana zero modes is reported in Appendix A, where we extend the method to extremely long-range regimes and discuss the limit of validity.}

\section{Correlation functions}
\begin{figure}
[!h]
\centering
\includegraphics[width=0.6\columnwidth]{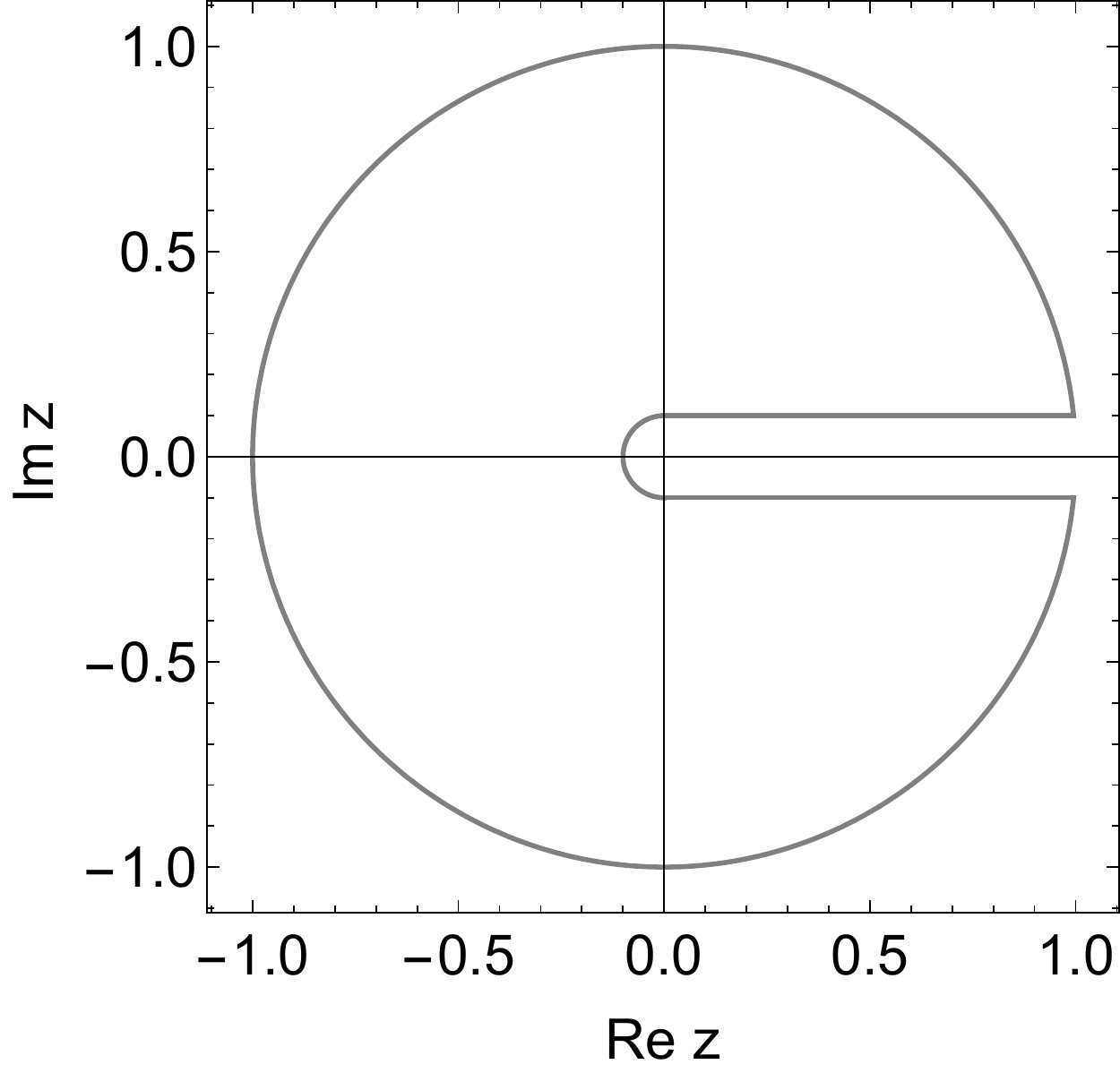}
\caption{Schematic representation of the path of integration in Eqs. (\ref{Cr0}), (\ref{Fr0}), for the correlation functions $C_{R0}$ and $F_{R0}$. The integrand has a branch cut on the positive real semi-axis.
}
\label{fig:path}
\end{figure}

{\change Let us proceed by calculating the correlation functions of the model which will play a fundamental role in our discussion.}
In the thermodynamic limit, the correlation functions 
\begin{eqnarray}
\label{cf}
C_{ij}=\langle a^\dagger_i a_j\rangle, \;\;F_{ij}=\langle a^\dagger_i a^\dagger_j \rangle
\end{eqnarray}
which depend on the relative distance $R$ between $i$ and $j$, read
\begin{eqnarray}
C_{R0} &=& \frac{\delta_{R,0}}{2}+\frac{1}{2 \pi}\int_0^\pi dk \cos(kR) \frac{\mu+w\, g(k) }{\epsilon_k}\,,\\
F_{R0} &=& -\frac{1}{2\pi}\int_0^\pi dk \sin(kR) \frac{\Delta\, f(k)}{\epsilon_k}\,.
\end{eqnarray}
We can calculate them by writing the integrals in the complex plane as follows
{\change 
\begin{eqnarray}
\label{Cr0}
C_{R0} &=& \frac{\delta_{R,0}}{2}+\frac{1}{4\pi} \textrm{Im}\oint_{|z|=1}\hspace{-0.15cm} dz \frac{\big(\mu +w\,\tilde g(z)
\big)\, z^{R-1}}{ \sqrt{\big(\mu+w \,\tilde g(z)\big)^2 -\big(\Delta \,\tilde f(z)\big)^2}}\,,\; \\
F_{R0} &=& \frac{1}{4\pi } \textrm{Im}\oint_{|z|=1}dz \frac{\Delta \,\tilde f(z)\, z^{R-1}}{ \sqrt{\big(\mu+w\,\tilde g(z)\big)^2 -\big(\Delta  \,\tilde f(z)\big)^2}} \,,
\label{Fr0}
\end{eqnarray}
where the path of integration is drawn in Fig.~\ref{fig:path}. The functions $\tilde g(z)$ and $\tilde f(z)$ are defined as 
$\tilde g(z)=(Li_\alpha(z)+Li_\alpha(1/z))$ and $\tilde f(z)=(Li_\beta(z)-Li_\beta(1/z))$. 
}
By using the residue theorem, a pole $z_0$ of the integrand  which is inside the unit circle gives an exponential decay $z_0^R$, conversely the brunch cut of the polylogarithm  gives an algebraic decay, which goes as
{\change 
\begin{eqnarray}
\label{CR0}
C_{R0} &\sim& \textrm{Im} \int_0^1 dx \frac{\big(\mu+ w\,\tilde g(x)
\big)\,x^{R-1}}{\sqrt{\big(\mu+w\,\tilde g(x)
\big)^2-\big(\Delta \,\tilde f(x)
\big)^2}}\,,\\
F_{R0} &\sim& \textrm{Im} \int_0^1 dx \frac{\Delta \,\tilde f(x)
\,x^{R-1}}{\sqrt{\big(\mu+w \,\tilde g(x)
\big)^2-\big(\Delta  \,\tilde f(x)
\big)^2}}\,.
\label{FR0}
\end{eqnarray}
where with $\tilde g(x)$ and $\tilde f(x)$ we mean 

\begin{eqnarray}
\tilde g(x)=Li_\alpha(x+i 0^+)+Li_\alpha(1/x-i 0^+)\\
\tilde f(x)=Li_\beta(x+i 0^+)-Li_\beta(1/x-i 0^+)
\end{eqnarray}
In the asymptotic limit, for very long distances, $R\to \infty$, }
$x^R$ is non zero only near one, thus we approximate the integrand with its expression near to one. 
{\change In this limit we can expand the polylogarithms around $|z|=1$ using, for non-integer $s$, the relation
\begin{equation}
Li_{s}(z)=\Gamma(1-s)(-\ln z)^{s-1}+\sum_{n=0}^\infty\frac{\zeta(s-n)}{n!}(\ln z)^n
\end{equation}
where $\Gamma(s)$ is the Gamma function and $\zeta(s)$ is Riemann zeta function. 
In addition we will need also $Li_1(z)=-\ln(1-z)$.
Inserting this expansion in Eqs.~(\ref{CR0}), (\ref{FR0}), taking the imaginary part and performing the integrals we can derive the asymptotic decays of  the correlators.}
For instance, if $\beta < \min\{1,\alpha\}$, 
the term 
{\change $\Delta \tilde f(x)$ dominates in the denominator of Eq.~(\ref{FR0})}, so that we get $F_{R0} \sim \int_0^1 x^{R-1} dx = R^{-1}$. Similarly, we can calculate the asymptotic formulas for all the correlation functions in all the situations, 
{\change getting 
\begin{eqnarray}
C_{R0}\sim 1/R^{a}\\
F_{R0}\sim 1/R^{b}
\end{eqnarray}
where the decay exponents $a$ and $b$ depend on $\alpha$ and $\beta$ as reported in Table~\ref{table: decay}.
We notice that for any $\alpha>1$, the asymptotic decays of the correlation functions behave exactly like in the case of purely long-range pairing ($\alpha\rightarrow \infty$) \cite{vodola14}. On the contrary the results for $\alpha<1$, reported in the second table of Table~\ref{table: decay}, have never been derived so far. 
}
\begin{table}
[!h]
\caption{The exponents $a$ and $b$ of the algebraic decay of the correlation functions $C_{R0}\sim R^{-a}$ and $F_{R0}\sim R^{-b}$, for $\alpha>1$ and $\alpha<1$. Notice that for $\alpha>1$ and $\mu=-2 Li_\alpha(1)$ the exponents can be different, for instance we get $b=1$ for any $\beta$.}
\vspace{0.2cm}
\begin{tabular}{|c||c|c|c|c|}
\hline
$\alpha>1$ &$\beta<1$     & $\beta=1$ & $1<\beta<2$ & $\beta>2$  \\
\hline\hline
$a$ & $2-\beta$     & $2$ & $2\beta-1$ & $\beta+1$\\
\hline
$b$& $1$& $1$ & $\beta$ & $\beta$\\
\hline
\end{tabular}
\\
\vspace{0.2cm}
\begin{tabular}{|c||c|c|c|c|c|}
\hline$\alpha<1$ &$\beta<\alpha$     & $\beta=\alpha$ & $\alpha<\beta<2$ & $\beta=\alpha+1$ & $\beta>2$  \\
\hline\hline
$a$ & $1-\beta+\alpha$     & $2-\beta$ & $1+2\beta-2\alpha$ & $5-\beta$ & $5-2\alpha$\\
\hline
$b$& $1$& $1$ & $1+\beta-\alpha$ & $4-\beta$ & $3-\alpha$\\
\hline
\end{tabular}
\label{table: decay}
\end{table}
\section{Long-range entanglement}
We first recall 
{\change some basics about the long-range entanglement in a gapped quantum phase. In general,} a state $\ket{\Phi}$ is short-range entangled if and only if there is a quantum circuit with a finite depth $U^M_{circ} = U^{(M)}_m \cdots U^{(2)}_m  U^{(1)}_m $ where $U^{(i)}_m $ is a piecewise local unitary with range $m$,  such that $\ket{\Phi} = U^M_{circ} \ket{\Phi_0}$, where $\ket{\Phi_0}$ is a product state (see Ref.~\cite{chen10} for details). In this case, any site can be correlated only with the sites in a neighborhood smaller than $J=2(Mm-M+1)$. As a result, if we divide the chain into three blocks $A$, $B$ and $C$, with $C$ between $A$ and $B$ whose size $\ell_{AB} \geq J$, the reduced matrix for the subsystem $A\cup B$ is $\rho_{A\cup B}=\rho_{A}\otimes \rho_B$. This implies that if there is short-range entanglement, then there are no correlations between arbitrary parties $A$ and $B$ if their spatial separation $\ell_{AB}$ is large enough.

\begin{figure}
[!h]
\centering
\includegraphics[width=0.9\columnwidth]{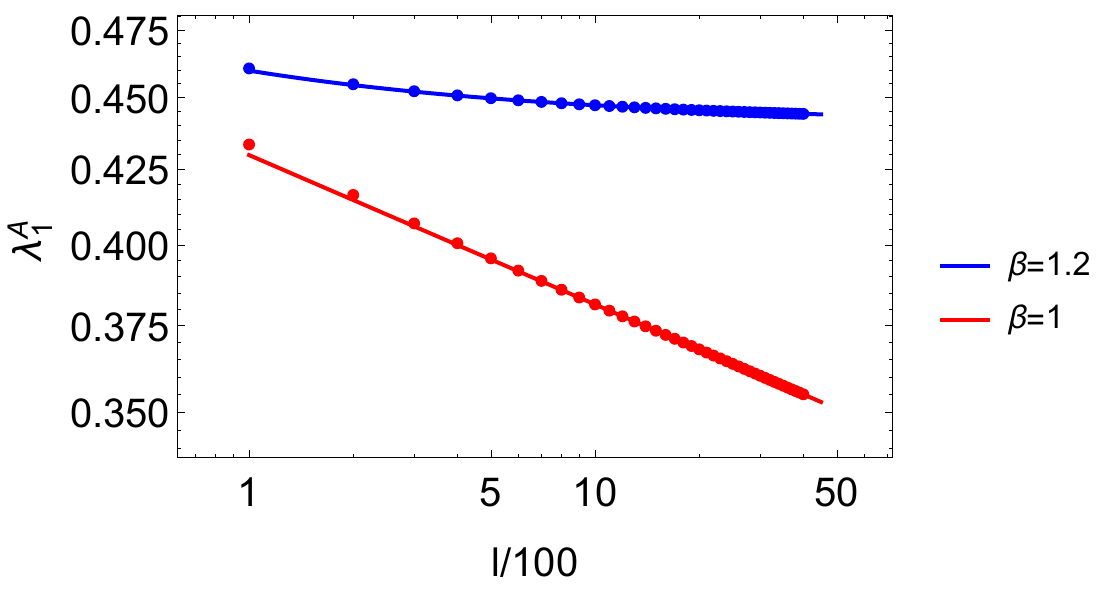}\\
\includegraphics[width=0.9\columnwidth]{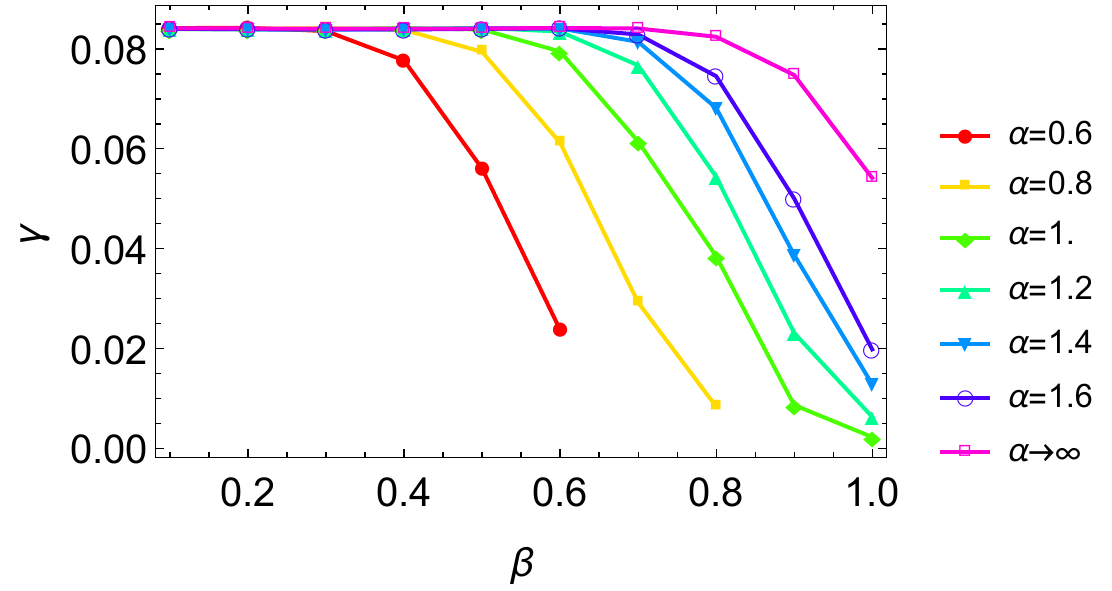}
\caption{(Top panel) Plot of $\lambda^A_1$, as a function of the block size $\ell$ for $\beta=1$ (red dots) and $\beta=1.2$ (blue dots). We put $w=\Delta=1$, $\mu=0$ and $\alpha\to \infty$. 
The blue line is the best fit with the function $c + c' \ell^{-\gamma}$, where $c\approx 0.441$, $c'\approx 0.17$ and $\gamma\approx0.471$, while the red line is the best fit with the function $c' \ell^{-\gamma}$, where $c'\approx 0.546$ and $\gamma\approx 0.052$ (the best fits have been performed for $\ell\geq 1000$).
(Bottom panel) Plot of $\gamma=c_{\textrm{eff}}/6$ as a function of $\beta$ for different values of $\alpha$ (obtained by best fitting in the interval $\ell\in[1000,2000]$). For small $\beta$ and such that $\beta<\min\{1,\alpha\}$, the value of $\gamma$ approaches $0.083\approx1/12$.
}
\label{fig:lambda1_gamma}
\end{figure}

\begin{figure}
[!h]
\centering
\includegraphics[width=0.75\columnwidth]{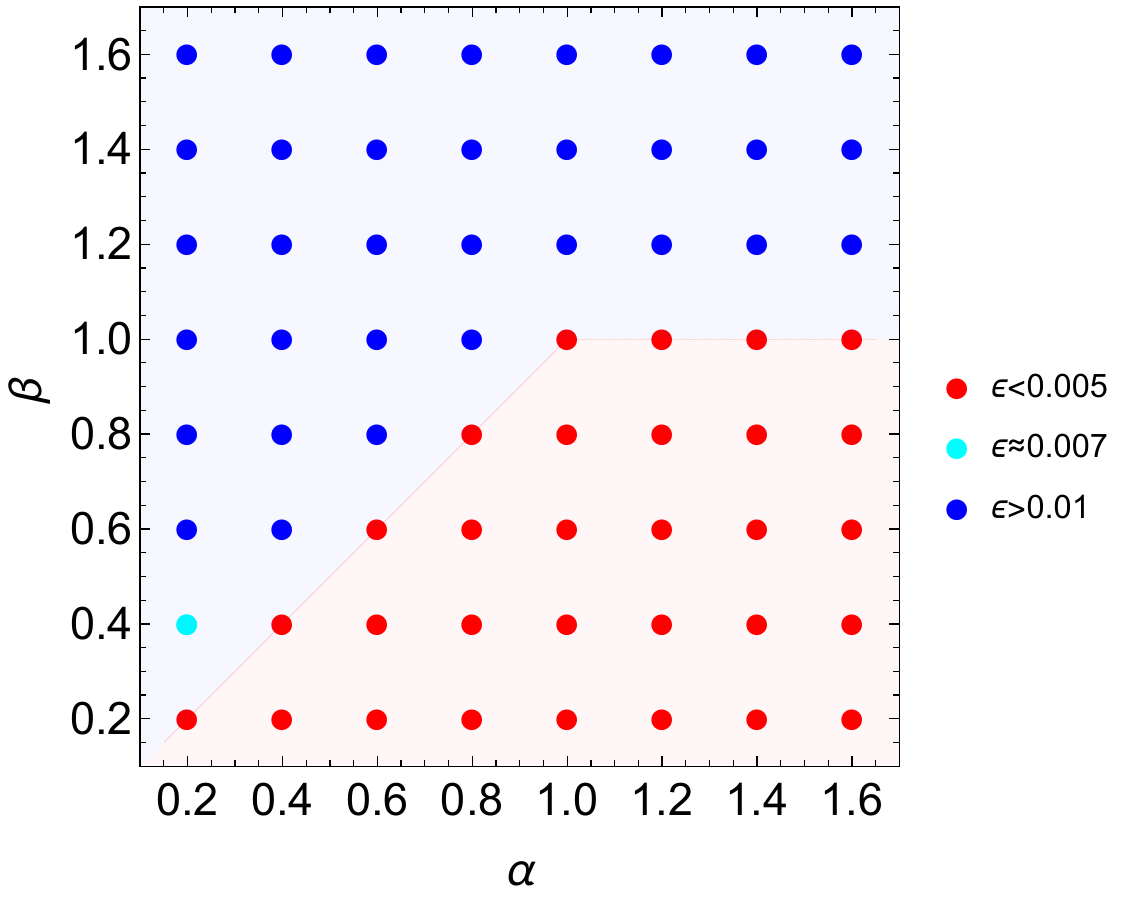}
\caption{Plot of $\epsilon\equiv (p_{\ln \ell,\ln \lambda^A_1}+1)$, namely the Pearson correlation coefficient plus one, in the $\alpha$-$\beta$ plane. We put $w=\Delta=1$, $\mu=0$. 
To calculate $\epsilon$, we considered the interval $\ell\in [300,2000]$ by changing $\ell$ by steps of $100$, and we calculate $\lambda^A_1$ at the values $\alpha=0.2 + 0.2\times n$ and $\beta=0.2 + 0.2\times n$ with $n=0,1,\dots,7$. We verified in this way that, for $\beta\le \min\{1,\alpha\}$, in the red region, $\lambda^A_1\sim \ell^{-\gamma}$, for large enough $\ell$. 
}
\label{fig:lambda1}
\end{figure}

{\change \subsection{Entanglement spectrum}}
Other useful quantities for characterizing the long-range entanglement are the entanglement spectrum and the entanglement entropy of a block $A$ with $\ell$ sites. We recall that the eigenvalues $\lambda^A_i$ (in non-increasing order) of the ground-state reduced density matrix $\rho_A$ of the block $A$, i.e., the square of the Schmidt coefficients, form the entanglement spectrum, and the entanglement entropy (the von Neumann entropy) can be expressed as $S_A(\ell)=-\sum_i \lambda^A_i \ln \lambda^A_i$. By considering short-range entanglement and the three blocks $A$, $B$ and $C$, one can show that (see, for instance, Ref.~\cite{perez-garcia07}) $\ket{\Phi} = \sum_{\alpha,\beta,\eta} {\cal A}_{\alpha\eta \beta} \ket{\psi^A_\alpha} \ket{\psi^C_\eta} \ket{\psi^B_\beta}$, with $1\leq \alpha,\beta,\eta \leq D$, where $D=2^J$ is the dimension of the Hilbert space of $C$. Then, there are maximum $D$ non-zero eigenvalues $\lambda^A_i$, and the entanglement entropy is bounded by $S_A \leq \ln D$, where the bound does not depend on the size $\ell$ of the block $A$.  
We note that if there is a finite number of non-zero $\lambda^A_i$ for any block $A$, then there is short-range entanglement (since the state is a matrix product state with finite dimensions~\cite{fidkowski11}).
To calculate the entanglement spectrum, as shown in Ref.~\cite{chung01}, we note that the reduced density matrix of a block $A$ can be expressed as 
\begin{equation}
\label{rho}
\rho_A =e^{-\sum_k \varepsilon_k^A f^\dagger_k f_k}/Z_A
\end{equation}
where $f_k$ are fermionic operators (linked to the original fermionic operators $a_i$ by a unitary transformation) and $\pm\tanh(\varepsilon^A_k/2)$ are the eigenvalues of the matrix
$\left(
      \begin{array}{cc}
        2 C_A -\mathbb{I} & 2 F_A \\
        - 2 F_A^* & -2C_A^* + \mathbb{I} \\
      \end{array}
    \right)\,$,
where $C_A$ and $F_A$ are correlation functions associated to the block $A$. 
Actually, $\tanh^2(\varepsilon^A_k/2)$ are the eigenvalues of $(2 C_A -I_A - 2 F_A)(2 C_A - I_A + 2 F_A)$.
Once obtained $\varepsilon^A_k$, we can write the $2^\ell$ eigenvalues of Eq.~(\ref{rho})
\begin{equation}
\lambda^A_i =  e^{-\sum_k \zeta^{(i)}_k \varepsilon^A_k}/Z_A,
\end{equation}
where $\zeta^{(i)}_k=0,1$ and $Z_A = \prod_k (1+e^{-\varepsilon^A_k})$. 
{\change It is worth noticing} that the long-range entanglement can be fully characterized by the greater eigenvalue $\lambda^A_1=1/Z_A$. If $\lambda^A_1\to 0$ when the block size $\ell$ tends to infinity, we get $S_A(\ell)\geq -\ln \lambda^A_1 \to \infty$, impling long-range entanglement.
If there is a number $N_A$ of finite $\varepsilon^A_k$, we get $N_A\to \infty$, if and only if $\lambda^A_1 \to 0$. 
If $\lambda^A_1 \to const$, then $N_A$ is finite so that there is a finite number of non-zero eigenvalues $\lambda^A_i$, implying short-range entanglement.
Actually, for our model, we find two distinct behaviors for $\lambda^A_1$: either $\lambda^A_1 \sim c + c' \ell^{-\gamma}$, or  $\lambda^A_1 \sim \ell^{-\gamma}$, for large enough $\ell$ (see Fig.~\ref{fig:lambda1_gamma}). 
To detect the long-range entanglement we looked at the Pearson correlation coefficient $p_{\ln \ell,\ln \lambda^A_1}$ between the variables $\ln \lambda^A_1$ and $\ln \ell$. This quantity tends to $-1$ 
{\change if  $\ln\lambda^A_1 \sim -\gamma \ln\ell +const$, i.e.,} 
in the presence of long-range entanglement. 
The Pearson correlation coefficient is defined as follows.
Given two sets of variables $\{x_1,\cdots, x_n\}$ and $\{y_1,\cdots, y_n\}$, we define the Pearson correlation coefficient the quantity $p_{x,y} = cov(x,y)/\sqrt{var(x)var(y)}$, where $cov$ is the covariance 
and $var$ is the variance.  
As shown in Fig.~\ref{fig:lambda1}, we observe long-range entanglement for $\beta\leq\min\{1,\alpha\}$. Actually, we also find that, for $\beta \le 1$, the exponent $\gamma$ appearing in the asymptotic bahavior for the entanglement spectrum, $\lambda_A\sim\ell^{-\gamma}$, saturates to the value $\gamma \simeq 0.083\approx 1/12$, (see Fig.~\ref{fig:lambda1_gamma}, bottom panel) for $\beta \lesssim \alpha$, extending the known result for the entanglement entropy for long-range paring \cite{vodola14, lepori17, ares}, also in the presence of both long-range hopping and pairing terms, getting
\begin{equation}
S_A=\frac{c_{\textrm{eff}}}{6}\log\ell
\end{equation}
where the effective central charge reads $c_{\textrm{eff}}= 6\gamma\approx 0.5$, for $\beta\le\min\{1,\alpha\}$.
{\change We have, therefore, extended the calculation for the entanglement entropy and the effective central charge, so far known only for long-range pairing \cite{vodola14,lepori17}, also in the presence of both long-range hopping and pairing.}

\begin{figure}
[!t]
\includegraphics[width=0.8\columnwidth]{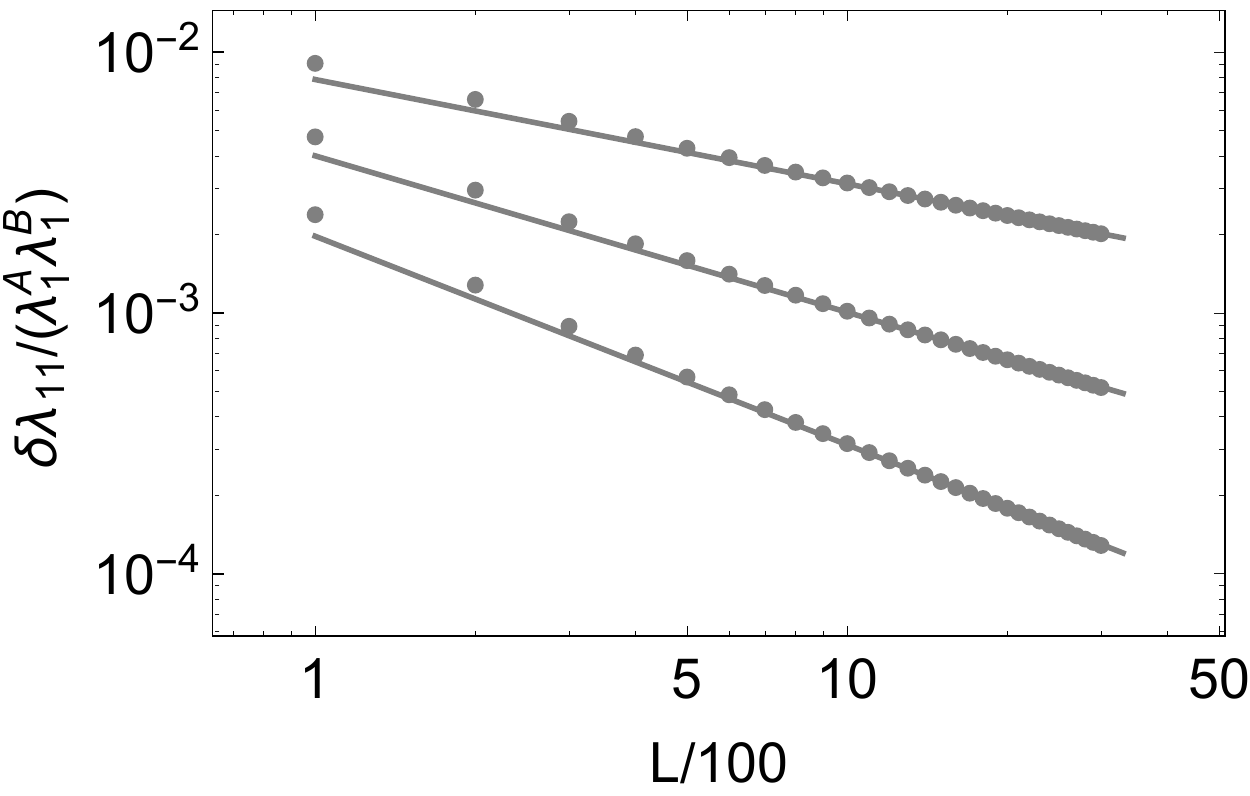}
\caption{Plot of $\delta \lambda_{11}/(\lambda^A_1\lambda^B_1)$ as a function of $L$, for different values of $\beta$, where $x=1/4$, $y=(1-2 x)/2$, $\alpha=2$, $w=\Delta=1$, $\mu=0$ and $\beta=1.2,1.3,1.4$ (from top to bottom). The solid lines are obtained by calculating the best fit 
with a a function proportional to $L^{2(1-b)}$, 
in the interval $L\in [1000,3000]$, by changing $L$ by steps of $100$.
}
\label{fig:hyp}
\end{figure}

\begin{figure}
[!h]
\includegraphics[width=0.8\columnwidth]{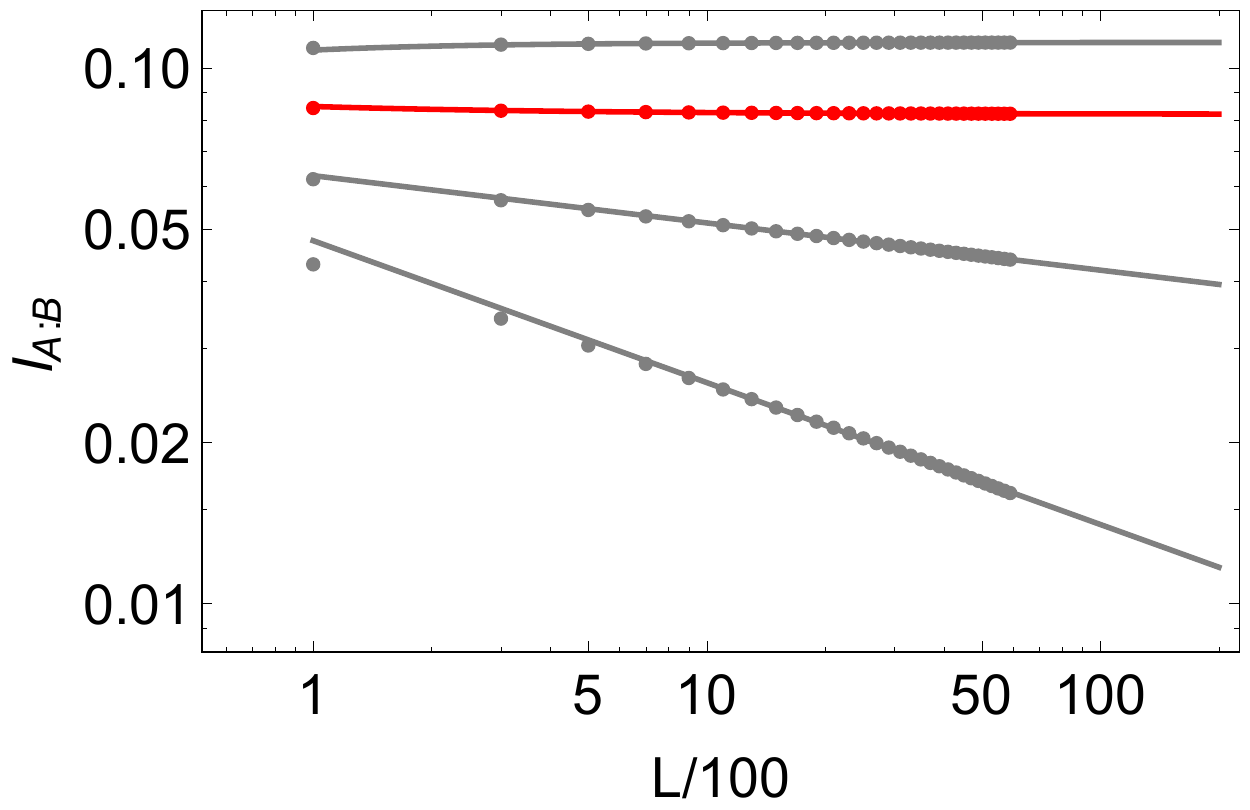}
\caption{Plot of $I_{A:B}$ as a function of $L$, for different values of $\beta$, where $x=1/4$, $y=(1-2 x)/2$, $\alpha\to \infty$, $w=\Delta=1$, $\mu=0$ and $\beta=0.9,1,1.1,1.2$ (from top to bottom). The value $\beta=1$ corresponds to the red line. The solid lines are obtained by performing an extrapolation, calculating the best fit in the interval $L\in [1000,6000]$, by changing $L$ by steps of $100$, with a function $a' L^{-\gamma'} + b'$ for $\beta\leq 1$ and $a' L^{-\gamma'}$ for $\beta>1$.
}
\label{fig:mi}
\end{figure}

{\change \subsection{Mutual information}}
It is worth observing that the presence of long-range entanglement is related to the decay of correlation functions.
{\change What we are going to see is that the algebraic decays of the correlation functions imply a peculiar long-range mutual information shared by two disjoint and infinitely distant regions.}
To explain this relation, we consider a closed chain of $L$ sites and two blocks $A$ and $B$ of sizes $L_A=L_B=\ell=x L$, separated by a distance $\ell_{AB} = y L$, with $x$ and $y$ some fraction of $L$.
{\change The plan is to calculate the reduced density matrix $\rho_{A\cup B}$ of the system made by the two blocks together with the reduced density matrices $\rho_{A}$ and $\rho_B$ of the two subsystems. 
To characterize the correlations between the two blocks $A$ and $B$, we consider the mutual information defined as 
\begin{equation}
\label{eq. mu in}
I_{A:B}=S_A + S_B -S_{A\cup B},
\end{equation}
where $S_{A\cup B}$ is the von Neumann entropy of the subsystem made by $A$ and $B$, while $S_A$ and $S_B$ the von Neumann entropy of the two blocks.
Quite in general, if we know the eigenvalues of the reduced density matrices
\begin{eqnarray*}
&&\rho_{A\cup B} \ket{\lambda^{AB}_{ij}} = \lambda^{AB}_{ij}\ket{\lambda^{AB}_{ij}}\\
&&\rho_A \ket{\lambda^A_i} = \lambda^A_i \ket{\lambda^A_i}\\
&&\rho_B \ket{\lambda^B_i} = \lambda^B_i \ket{\lambda^B_i}
\end{eqnarray*}
with $\lambda^{AB}_{ij}=\lambda^A_i \lambda^B_j + \delta \lambda_{ij}$, and $\delta \lambda_{ij}$ small enough, we can write
\begin{equation}\label{eq. mi}
I_{A:B} \sim \sum_{ij} \delta \lambda_{ij} \ln(\lambda^A_i\lambda^B_j) + \sum_{ij} \frac{(\delta \lambda_{ij})^2}{2 \lambda^A_i\lambda^B_j}\,.
\end{equation}
Let us assume that the correlation function $F_{R0}\sim R^{-b}$ dominates over $C_{R0}$ for large $R$, 
as one can easily check looking at Table~\ref{table: decay}. Under the following scaling hypothesis
\begin{equation}
\delta \lambda_{ij} \sim  \lambda^A_i \lambda^B_j \,L^{2(1-b)}
\end{equation}
which has been verified numerically, see Fig.~\ref{fig:hyp}, 
reminding that, according to Table~\ref{table: decay}, $b\ge 1$, we get the following scaling law for the mutual information 
\begin{equation}
\label{eq.mi_scal}
I_{A:B} \sim L^{2(1-b)}
\end{equation}
Analytical arguments supporting this result and the scaling hypothesis are gathered in Appendix B. Our result in Eq.~(\ref{eq.mi_scal}) is nicely verified numerically, as shown in Fig.~\ref{fig:mi}.
}
In particular, in the long-range regime, as shown in Fig.~\ref{fig:mi}, $I_{A:B}$ saturates to a constant value as expected, {\change in agreement with the decay exponent $b=1$, signature of a bulk long-range entanglement}, otherwise it vanishes as a power law, upon increasing $L$. 
As a remark, we note that if $C_{R0}$ have dominated over $F_{R0}$ we would have got $a$ instead of $b$ in Eq.~(\ref{eq.mi_scal}) but this situation never occurs, according to the decay exponents reported in Table~\ref{table: decay}. 

It is worth mentioning that the non-vanishing long-range mutual information implies that the disconnected entanglement entropy, 
$S_D=S_A + S_B -S_{A\cup B}-S_{A\cap B}$,  introduced in Ref.~\cite{zeng,magnifico} as a generalization of the so-called topological entanglement entropy \cite{kitaev06}, is an entanglement signature for symmetry-protected topological phases and sensitive to long-range entanglement between edges \cite{micallo}. On the other hand if the ground state is short-range entangled, for $A$ and $B$ two simply connected partitions of the chain separated by a large distance, then $\rho_{A\cup B}\sim \rho_A\otimes\rho_B$. In that case the mutual information $I_{A:B}$ of two disjoint and distant partitions is zero, as well as $S_D$. 
However, while for short-range interactions $S_D$ can be used as an order parameter for the topological phases, for long-rage pairing, the long-range interactions leads to the generation of long-range entanglement in the bulk states, as shown in Ref.~\cite{mondal}. In this case $S_D$ turns out to be finite in a wider range of chemical potential. In other words, the long-range couplings induces a sort of long-range entanglement in the bulk defined as the lack of a short-range one. 

Finally, let us give an 
analytical argument to explain the behavior reported in Eq.~(\ref{eq.mi_scal}), 
as obtained numerically.  Defining the matrices
\begin{equation}
\label{C}
{\cal C}_{AB}=\left(
\begin{array}{cc}
        \mathbb{I}- C_{AB} & F_{AB}^\dagger \\
        F_{AB} & C_{AB}\\
      \end{array}
\right)
\end{equation}
where $C_{AB}$ and $F_{AB}$ are matrices in real space whose elements, $C_{ij}$, $F_{ij}$ are the two-point correlation functions, Eq.~(\ref{cf}), where $i\in A$, $j\in B$ (where $A$ and $B$ can or cannot coincide). \\
For a connected subsystem, $A=B$, of size $\ell$, the eigenvalues of ${\cal C}_{AA}$, defined as in Eq.~(\ref{C}), are $n^{A}_{k +}=\frac{1}{1+e^{\varepsilon^{A}_k}}$ and $n^{A}_{k-}=\frac{1}{1+e^{-\varepsilon^{A}_k}}=1-n^{A}_{k+}$, where $\varepsilon^{A}_k$ are the eigenvalues of the effective Hamiltonian of the reduced density matrix, Eq.~(\ref{rho}), 
which can be easily written as 
{\change 
\begin{equation}
\rho_A=\otimes_{k=1}^{\ell}
\left(
\begin{array}{cc}
        n_{k+}^{A} &0 \\
        0 & n_{k-}^{A} \\
      \end{array}
\right).
\end{equation}}
As a result, the corresponding entanglement entropy 
{\change reads 
\begin{equation}
\label{SA}
S_A=-\sum_{k=1}^\ell \left(n^{A}_{k+}\log n^{A}_{k+}+n^{A}_{k-}\log n^{A}_{k-}\right).
\end{equation}}
Let us now consider a subsystem $A\cup B$, made by two disjoint segments, $A$ and $B$, and define
\begin{equation}
\label{C_{AB}}
{\cal C}_{A\cup B}=\left(
\begin{array}{cc}
        {\cal C}_{AA} & {\cal C}_{AB} \\
        {\cal C}_{BA}  & {\cal C}_{BB} \\
      \end{array}
\right)
\end{equation}
where the four elements are the correlation matrices given by Eq.~(\ref{C}). In order to calculate $S_{A\cup B}$ we have to find the eigenvalues 
$n^{A\cup B}$ of Eq.~(\ref{C_{AB}}) solving the following equation

\begin{eqnarray}
\label{det}
\nonumber
\det \big({\cal C}_{BB}-{\cal C}_{BA}({\cal C}_{AA}-n^{A\cup B}\mathbb{I})^{-1}{\cal C}_{AB}-n^{A\cup B}\mathbb{I}\big)&&\\
\times 
\det \big({\cal C}_{AA}-n^{A\cup B}\mathbb{I}\big)&&
\,=0
\end{eqnarray}
{\change We can, therefore, calculate the entanglement entropy 
\begin{equation}
\label{SAUB}
S_{A\cup B}= -\sum_{q=1}^{2\ell}\left(n^{A\cup B}_{q+}\log n^{A\cup B}_{q+}+n^{A\cup B}_{q-}\log n^{A\cup B}_{q-}\right).
\end{equation}
So far everything is exact and represents an alternative route with respect to that describe previously to calculate the entropies. 
Let us assume, for simplicity, that} ${\cal C}_{AA}={\cal C}_{BB}$ and ${\cal C}_{AB}={\cal C}_{BA}$, all $2\ell\times 2\ell$ matrices. Eq.~(\ref{det}), then, reduces to
\begin{equation}
\det \big({\cal C}_{AA}-{\cal C}_{AB}-n^{A\cup B}\mathbb{I}\big)\,\det \big({\cal C}_{AA}+{\cal C}_{AB}-n^{A\cup B}\mathbb{I}\big)=0
\end{equation}
valid also for non-commuting ${\cal C}_{AA}$ and ${\cal C}_{AB}$. We reduce the problem to simply finding the eigenvalues of $({\cal C}_{AA}\pm {\cal C}_{AB})$.\\ 
Let us now define, for simplicity, $q=(k,\pm)$, which can be enumerated from $1$ to $2\ell$, and 
the unitary transformation $U$ which diagonalizes ${\cal C}_{AA}$, namely $U^\dagger {\cal C}_{AA}U={\cal D}_A$ where ${\cal D}_A$ is a $2\ell$ diagonal matrix whose non-vanishing elements are $n^{A}_{q}$. 
Let us suppose that the eigenvalues $n^A_q$ are distinct. 
If ${\cal C}_{AB}$ can be considered as a perturbation of ${\cal C}_{AA}$, 
the eigenvalues of Eq.~(\ref{C_{AB}}) are, at first approximation, 
\begin{equation}
\label{nAUB}
n^{A\cup B}_{q\pm}=n^{A}_{q}\pm\delta n_q\simeq n^{A}_{q}\pm (U^\dagger {\cal C}_{AB}U)_{qq}
\end{equation}
{\change 
We point out that if $ {\cal C}_{AB} \neq { \cal C}_{BA}$, in order to get  $\delta n_q$, we have  to replace ${\cal C}_{AB}$ with $({\cal C}_{AB} + { \cal C}_{BA})/2$ (see Appendix C).} 
We notice that in ${\cal C}_{AA}$ the matrices $C_{AA}$ and $F_{AA}$ (previously called $C_A$ and $F_A$) are symmetric Toeplitz matrices (which almost commute with a matrix $J$ where all the elements are equal to one, if boundary terms can be neglected). In ${\cal C}_{AB}$, the matrices $C_{AB}$ and $F_{AB}$, for very large distances $\ell_{AB}$ can be approximated as $C_{AB}\sim\ell_{AB}^{-a} \,J$ and $F_{AB}\sim  \ell_{AB}^{-b}\,J$. 
For $b<a$, we have $F_{AB}$ dominating over $C_{AB}$ and we get
$\delta n_q\simeq \ell_{AB}^{-b} |\sum_{i=1}^{2\ell} U_{iq}|^2\sim \ell_{AB}^{-b}\ell^\eta$ 
where $0\le \eta\le 1$, depending on how much $U$ is a sparse matrix. 
{\change 
Inserting Eq.~(\ref{nAUB}) in Eq.~(\ref{SAUB}) and using Eq.~(\ref{SA}) we get}
\begin{equation}
S_{A\cup B}\simeq 2S_A
-2\sum_{q=1}^{2\ell}\frac{(\delta n_q)^2}{n^A_q}
\end{equation}
where in the first term we have exactly $S_A=\sum_{q=1}^{2\ell} n^{A}_{q}\log n^{A}_{q}$ while the second term is the mutual information at first approximation, which reads
$I_{A:B} \sim \ell_{AB}^{-2b}\ell^{2\eta}$. If either the size of the blocks and their distance are extensive, i.e. $\ell\propto L$ and $\ell_{AB}\propto L$, we have $I_{A:B} \sim L^{2(\eta-b)}$. Since $\eta$ is bounded, $\eta\le 1$, the mutual information surely vanishes for $b>1$.  This result is consistent with the leading term in Eq.~(\ref{eq. mi scal}), with $\delta=2$.

We argue that $\eta\simeq 1$. The reasoning goes as follows. Let us consider, for simplicity, the normal (non-superconducting) case, so that we have 
simply ${\cal C}_{AA}=C_{AA}$, which is a symmetric Toeplitz matrix, and ${\cal C}_{AB}=C_{AB}\sim \ell^{-a}J$. 
Supposing that $\ell$ is very large so that $C_{AA}$ can be confused with a matrix with periodic boundary condition, making, therefore, an error on the boundary terms which might become negligible for large sizes.  Under this assumption the unitary transformation $U$ is the Fourier transform $U_{jk}\simeq e^{ikj}/\sqrt{\ell}$ and, therefore, $\delta n_k\simeq \delta_{k0}\ell_{AB}^{-a}\ell$. As a result, $S_{A\cup B}\simeq 2S(A)-2\frac{(\delta n_0)^2}{n^A_0} $, namely $I_{A:B}\sim \ell_{AB}^{-2a}\ell^2$, meaning that $\eta=1$. 

\section{Dynamical properties}
We {\change complete} our discussion by focusing on {\change some} dynamical features, driven either by a sudden quench and by an adiabatic evolution. {\change In particular we are going to show that dynamical quantum phase transitions can occur even without crossing any phase boundary after a sudden quench from long to a short-range regions of the phase diagram. Looking at the universal adiabatic dynamics, instead, we will show how the scaling of the density of defects generated after crossing a quantum critical point, is related to the topological healing length, rather than the correlation length. This observation clarifies also the violation of the Kibble-Zurek mechanism for $1<\alpha<\min(2,\beta)$ found in 
Ref.~\cite{defenu19}}.

{\change \subsection{Sudden quench: dynamical quantum phase transitions}}
We start considering the time evolution generated by a sudden quench of the Hamiltonian, i.e., at the initial time $t_i=0$ the initial state of the system $\ket{\psi(t_{i})}$ is the ground-state of the initial Hamiltonian $H$, and then the Hamiltonian is suddenly changed to $H'$ (with apostrophized parameters, e.g., $\mu'$), generating the time evolution. Dynamical quantum phase transitions occur at the Fisher times, when the so-called Loschmidt amplitude, defined as $G(t)=\bra{\psi(t_{i})}e^{-iH'(t-t_i)}\ket{\psi(t_{i})}$, and in the thermodynamic limit, the free energy density $-\ln G(t)/L$ is non-analytic~\cite{heyl13}. 
{\change 
This quantity measure the probability amplitude for the system to return, during its time evolution, to the initial state after the Hamiltonian $H$ has been changed to $H'$.} 
As shown in Ref.~\cite{vajna15}, the presence of dynamical phase transitions can be related to the topological features of the quantum phases. In general, 
{\change in the thermodynamic limit, the Loschmidt amplitude reads
\begin{equation}
G(t) = \lim_{L\to \infty} \exp \left(\frac{L}{\pi} \int_0^\pi \ln(\cos(\epsilon'_k t) + i \hat d_k \cdot \hat d'_k \sin(\epsilon'_k t) )dk\right)\,,
\end{equation}
and thus} 
there are dynamical phase transitions if there is at least a  $k$ such that $\vec d'_k \cdot \vec d_k=0$, and this is the case if we cross a topological quantum critical point with the quench. However, for our model, there are   dynamical phase transitions also if we cross the boundary between short-range and long-range entanglement.  To prove it, we note that in the thermodynamic limit we get $\vec d_\pi = -\sign(\mu+w g(\pi)) \hat e_3$, 
{\change and, 
\begin{eqnarray}
&&\vec d_0 = \sign(\Delta) \hat e_2\,,  \;\textrm{if} \;\;\beta < \min\{1,\alpha\}\\ 
&&\vec d_0 = -\sign(\mu+w g(0)) \hat e_3\,, \;\textrm{if} \;\;\beta > \min\{1,\alpha\}.
\end{eqnarray}
Thus, if the quench is from a long-range region to a short-range one, we get $\vec d_0 \cdot \vec d'_0=0$, so that there are dynamical phase transitions at the Fisher times 
\begin{equation}
t_n = (n+1/2)\pi/(\epsilon'_0).
\end{equation}
Vice versa, if $\beta'<\min\{1,\alpha'\}$, $\epsilon'_k $ diverges  as $k\to 0$, and $t_n$'s are dense in $(0,\infty)$, so that the non-analytic behavior occurs at any time, i.e., the free energy density is nowhere analytic in the complex plane.
}

{\change \subsection{Slow quench: Kibble-Zurek mechanism}}
We conclude our discussion by investigating how the density of adiabatic excitations, generated by crossing linearly in time a quantum critical point, 
(see for instance Refs.~\cite{zurek85}, ~\cite{Dziarmaga05}) 
is related to the topological features of the quantum phase transition. 
{\change We consider a chemical potential which changes linearly in time as $\mu(t)=t/\tau_Q$ for $t\in(-\infty,0)$, such that we cross the quantum critical point $\mu_c=-w g(0)$ for $\beta>1$ and $\alpha>1$. 
It has been shown in Ref.~\cite{defenu19} that, for $1<\beta <2$, the excitation density $n$ (see Appendix D for more details) scales as 
\begin{equation}
n\sim \tau_Q^{1/(2-2\beta)}
\end{equation}
    while for $\beta>2$, $n \sim {\tau_Q^{-1/2}}$. According to the Kibble-Zurek mechanism~\cite{zurek85}, this quantity is related to the universal exponents $\nu$ (related to the closure of the gap) and $z$ (the so-called dynamical exponent) as 
    \begin{equation}
    \label{KZm}
    n \sim \tau_Q^{-\nu/(1+z \nu)}.
    \end{equation}
    }
It is interesting to {\change compare the values of the exponent $\nu$, related to the scaling of the correlation length $\xi$, to the ones} of the characteristic {\change healing length $\xi_t$ close} to the topological phase transition, which is defined by considering the winding number
\begin{equation}
W = \frac{1}{2\pi} \int_{-\pi}^\pi \partial_k \theta_k dk
\end{equation}
where $\theta_k = \arcsin(\Delta f(k)/(\epsilon_k))$, such that (see, e.g., Ref.~\cite{cheng17})
\begin{equation}\label{eq xi}
{\change \int_{k_c-\xi_t^{-1}}^{k_c+\xi_t^{-1}} \partial_k \theta_{k} dk = O(1)}
\end{equation}
where $\partial_k \theta_{k}$ can be approximated by the Taylor expansion {\change around  $k_c$, the point where the gap closes.} 
For $\mu_c=-w g(0)$ {\change ($k_c=0$)} and $1<\beta<2$, {\change and always $\alpha>1$,} we get 
\begin{equation}
{\change  \partial_k \theta_k \sim {|k|^{\beta-2}}/{|\mu-\mu_c| }}
\end{equation}
as $k \to 0$, thus by substituting in Eq.~\eqref{eq xi}, considering the principal value of the integral, we get 
 {\change
 \begin{equation}
 \xi_t \sim |\mu -\mu_c|^{1/(1-\beta)}.
 \end{equation}
 so that we can define $\nu_t=1/(\beta-1)$.  For $\beta>2$ and $\alpha>1$ we get, instead, $\partial_k \theta_k \sim 1/ |\mu-\mu_c|+O(k^2)$, getting $\nu_t=1$. 
This results for $\nu_t$ are in agreement with the values of $\nu$ 
only for $\alpha>\min(2,\beta)$, 
while, for $\alpha<\min(2,\beta)$, we obtain $\nu_t\neq \nu$, as shown in Table \ref{table2}. In the latter case there is a violation of the Kibble-Zurek mechanism, called $\beta$-dynamics \cite{defenu19}. In all the cases $z\nu=1$.}\\
\begin{table}[h!]
\caption{{\change $\nu$ and $\nu_t$ for different values of $\alpha$ and $\beta$. We consider $\alpha>1$ and $\beta>1$.}}
{\change \begin{tabular}{|c||c|c|c|c|}
\hline
&$\alpha>2$, $\,\beta>2$     & $\alpha>\beta$, $\,\beta<2$ & $\alpha<\beta$, $\,\beta<2$ &  $\alpha<2$, $\,\beta>2$\\
\hline\hline
$\nu$ & $1$     & $1/(\beta-1)$ & $1/(\alpha-1)$  & $1/(\alpha-1)$\\
\hline
$\nu_t$& $1$& $1/(\beta-1)$ & $1/(\beta-1)$ & $1$\\
\hline
\end{tabular}}
\label{table2}
\end{table}
{\change In conclusion we propose that the Kibble-Zurek mechanism for one-dimensional long-range systems, rather than been described by Eq.~(\ref{KZm}), has to be modified as follows
\begin{equation}
\label{KZmt}
n \sim \tau_Q^{-\nu_t/(1+z \nu)}.
\end{equation}

\bigskip
\noindent 
Finally, let us consider the case $\mu>0$, so that the gap closes at $k_c=\pi$. We get $\partial_k \theta_k \sim 1/ |\mu-\mu_c|+O((k-\pi)^2)$, 
with $\mu_c=-wg(\pi)$, therefore 
$\xi_t \sim |\mu -\mu_c|^{-1}$, like the correlation length $\xi$. In this last case $\nu=\nu_t=1$, getting 
simply $n \sim \tau_Q^{-1/2}$.

}

\section{Conclusions}
We performed a complete study of the correlation functions for the Kitaev model with both long-range hopping and pairing, {\change 
expected to describe experimental realizations of long-range topological superconductors \cite{perge,pawlak,ruby},} 
finding all the analytical expressions for their algebraic asymptotic decays (see Table \ref{table: decay}). 
Moreover we find that the critical-like behavior of the entanglement entropy 
can be extended also in the presence of long-range hopping, as far as $\beta<\min\{1,\alpha\}$. 
We investigated the condition for getting long-range mutual information shared by two extended disconnected regions. This quantity has to be finite in order to get a finite disconnected entanglement entropy which detects long-range entanglement entropy at the edges. We show that, deep in the long-range interacting regime, the mutual information between two generic segments is always finite even at infinite distances. This implies that the reduced density matrix of a composite subsystem cannot be factorized, getting a sort of long-range entanglement in the bulk. 
Looking at the time evolution generated by a quench between short-range and long-range entanglement regions, we showed that there are   dynamical quantum phase transitions, {\change also without crossing any phase boundary}. 
Finally, we 
showed that the Kibble-Zurek mechanism is related to a topological scale length {\change at the quantum critical point, and how it should be modified in the presence of long-range couplings.} 

\subsection*{Acknowledgements}
The authors acknowledge financial support from the project BIRD 2021 "Correlations, dynamics and topology in long-range quantum systems" of the Department of Physics and Astronomy, University of Padova.

\appendix
{\change 
\section{Majorana zero modes}
}
In the Majorana basis, $\lambda_k = a_k + a_k^\dagger$, $\lambda_k'=i a_k^\dagger - i a_k$, 
we can rewrite Eq.~(\ref{eq. hami k}) getting $H= i/4 \sum_k \Lambda_k^T X(k) \Lambda_{-k}$, with $\Lambda_k=(\lambda_k,\lambda'_{-k})^T$ and $X(k) = i(w g(k) + \mu)\tau_2 - \Delta f(k) \tau_0$, where $\tau_0$ is the identity matrix.

For short range interactions, the Majorana number~\cite{kitaev01} indicates the presence of edge modes (if it is minus one) for the case of an open chain, and it is equal to $\sign((\mu+g(0)w)(\mu+g(\pi)w))$. 
The proof is as follows: In order to evaluate the parity for $L$ odd for a closed chain, we note that
\begin{equation}
H= \frac{i}{4} \Lambda_0^T X(0) \Lambda_0+\frac{i}{4}\sum_{k>0}\left(
                                                       \begin{array}{cc}
                                                         \Lambda_k^T & \Lambda_{-k}^T \\
                                                       \end{array}
                                                     \right)
 \left(
                                                        \begin{array}{cc}
                                                           0 & X(k) \\
                                                          X(-k) & 0  \\
                                                        \end{array}
                                                      \right) \left(
                                                                \begin{array}{c}
                                                                  \Lambda_k \\
                                                                  \Lambda_{-k} \\
                                                                \end{array}
                                                              \right)
\end{equation}
We consider  $A=X(0)\oplus \left(\oplus_{k>0} \left(
                                                        \begin{array}{cc}
                                                          0 & X(k) \\
                                                          X(-k) & 0 \\
                                                        \end{array}
                                                      \right)\right)$, then the parity is the sign of the Pfaffian of $A$. Since $\det X(k) >0$, by using the properties of the Pfaffian it is easy to show that $\sign Pf A = \sign Pf X(0)=\sign(\mu+g(0)w)$.
Similarly, for $L$ even the parity is $\sign Pf X(0) Pf X(\pi)= \sign((\mu+g(0)w)(\mu+g(\pi)w))$, from which the expression of the Majorana number.

For an open chain, the quadratic Hamiltonian can be diagonalized with the general method reported in Ref.~\cite{lieb61}. By performing a numerical investigation, we find that in the region where the Majorana number is $-1$, there can be Majorana zero modes which can acquire a mass if $\beta$ is small enough while they are massless if $\beta>\min\{1,\alpha\}$, and their wave-functions become bi-localized at the edges, forming a 
non-local complex Dirac fermion.

The topological phase of the model, therefore, is characterized by Majorana zero modes. In particular at the symmetric point $\mu=0$, $\Delta=w$ and $\alpha=\beta$, the Majorana fermions $c_1$ and $c_{2L}$ are decoupled, {\change where $c_{2j-1}=a_j + a^\dagger_j$ and $c_{2j}=i a^\dagger_j - i a_j$}. By writing $H=\sum_{m,n}c_m {\cal H}_{mn}c_n$, where $i {\cal H}$ is the real and skew-symmetric matrix
\begin{eqnarray}
\nonumber {\cal H} &=& \sum_{i,j} \ket{i}\bra{j}\otimes {\cal H}_{i,j} = \sum_j \ket{j}\bra{j} \otimes  {\cal H}_0  \\
&& + \sum_j\sum_l  \ket{j} \bra{j+l} \otimes {\cal H}_l +  \ket{j+l}\bra{j}\otimes ({\cal H}_l)^\dagger\,,
\end{eqnarray}
where ${\cal H}_0$ and ${\cal H}_l$ are the matrices ${\cal H}_0 = \mu \tau_2 /4$ and ${\cal H}_l = w l^{-\alpha}\tau_2/4 + i \Delta l^{-\beta} \tau_1/4$, at the symmetric point  $\ket{1}$ and $\ket{2L}$ are eigenvectors of ${\cal H}$ with zero eigenvalue, where the nth component of $\ket{i}$ is $(\ket{i})_n=\delta_{n,i}$. In general, if there are {\change Majorana zero modes} then there are two eigenvectors $\ket{v_1}$ and $\ket{v_{2L}}$ of ${\cal H}$ with zero eigenvalue and localized at the edges, and their components give the wave-functions of the Majorana fermions. To derive a condition for the existence of {\change Majorana zero modes} and study the decay of their wave-functions, we define the projectors $P=\ket{1}\bra{1}+\ket{2L}\bra{2L}$ and $Q=I-P$, so that a generic matrix, in our case $\cal H$, can be written in a block form as
\begin{equation}
{\cal H}=\left(
    \begin{array}{cc}
      {\cal H}_P & {\cal H}_{PQ} \\
      {\cal H}_{QP} & {\cal H}_Q \\
    \end{array}
  \right)\,,
\end{equation}
where ${\cal H}_P$ is the block corresponding to the subspace of $P$, and so on. We consider the eigenvalue equation ${\cal H}\ket{v}=E\ket{v}$, from which we get $P({\cal H}-E)\ket{v}=0$, or equivalently $({\cal H}_P-E  \mathbb{I}_P)\ket{v_P}+{\cal H}_{PQ}\ket{v_Q}=0$, and $Q({\cal H}-E)\ket{v}=0$, i.e., ${\cal H}_{QP}\ket{v_P}+({\cal H}_Q-E \mathbb{I}_Q)\ket{v_Q}=0$. By combining these two equations, for $E=0$, we get that there are 
{\change Majorana zero modes} if the $2\times 2$ matrix
\begin{equation}
\Gamma_P = {\cal H}_{PQ} \frac{1}{{\cal H}_Q} {\cal H}_{QP}
\end{equation}
tends to zero as $L\to \infty$, where we have considered ${\cal H}_P \to 0$ in this limit, and ${\cal H}_Q$ non-singular. The wave-function of the Majorana fermion in the bulk is obtained by
\begin{equation}
\ket{v_Q} = - \frac{1}{{\cal H}_Q} {\cal H}_{QP} \ket{v_P}\,,
\end{equation}
where $\ket{v_P}$ is a  two-dimensional vector.

We note that the matrix ${\cal H}_{QP}$ has elements $({\cal H}_{QP})_{m, 1} = {\cal H}_{m+1, 1}$ and $({\cal H}_{QP})_{m, 2} = {\cal H}_{m+1, 2L}$, with $m=1,\cdots,2L-2$. 

It is easy to see that $({\cal H}_{QP})_{2j-1, 1} =i\mu/4 \delta_{j,1}+i/4(1-\delta_{j,1})(w/(j-1)^\alpha-\Delta/(j-1)^\beta)$, thus if $\alpha=\beta$ and $\Delta=w$, if $\ket{v_P}=(1,0)^T$, we get $(\ket{v_Q})_j \sim \mu ({\cal H}_Q^{-1})_{j,1}$ and $(\Gamma_{P})_{2,1}\sim \mu^2 ({\cal H}_Q^{-1})_{2L-2,1}$. Thus, since $({\cal H}_Q^{-1})_{2j-1,1}=0$, it is enough to study the asymptotic behavior of $({\cal H}_Q^{-1})_{2j,1}$. By performing a numerical investigation, we find that $({\cal H}_Q^{-1})_{2j,1}$ can decay for certain values of the parameters, then there are {\change Majorana zero modes} having wave functions with the same decay of $({\cal H}_Q^{-1})_{2j,1}$. For other values of the parameters, $({\cal H}_Q^{-1})_{2j,1}$ does not decay, so that $\Gamma_P\neq 0$ and there are no Majorana fermions.
To perform an analytic study, we consider periodic boundary conditions, and we change basis by defining the vectors $\ket{k}$ such that $\ket{j} = \sum_k e^{-i k j}\ket{k}/\sqrt{L}$, so that we get ${\cal H} = \sum_k \ket{k} \bra{k}\otimes {\cal H}_k$, where 
${\cal H}_k = \left( (\mu + w g(k))\tau_2 - \Delta f(k)\tau_1\right)/4$. The inverse  of the matrix ${\cal H}$ reads
\begin{equation}
{\cal H}^{-1} = -4 i \sum_k \ket{k}\bra{k}\otimes \left(
                         \begin{array}{cc}
                            0 & 1/X_+ \\
                           -1/X_- &  0 \\
                         \end{array}
                       \right) \,,
\end{equation}
where $X_\pm = \mu + w g(k)\pm i \Delta f(k)$. 
Then, for the block correspondent to the subspace of $Q$, we get 
\begin{eqnarray}
\label{eq. decay zero Majorana}
&&(({\cal H}^{-1})_Q)_{2j,1} = \\
&&\nonumber \frac{2}{\pi} \textrm{Im} \oint_{|z|=1} \frac{dz\, z^{j-1}}{\mu + w (Li_\alpha(z)+Li_\alpha(1/z))+ \Delta(Li_\beta(z)-Li_\beta(1/z))}\,.
\end{eqnarray}
In particular, the decay of $({\cal H}_Q^{-1})_{2j,1}$ can be approximated with the one of $(({\cal H}^{-1})_Q)_{2j,1}$ if the off-diagonal terms ${\cal H}_{PQ}$ and ${\cal H}_{QP}$ are negligible, i.e., if $\mu\approx 0$,  $w\approx \Delta$ and $\alpha\approx \beta$.
We note that for $\Delta=w$ and $\beta=\alpha$, the function $Li_\alpha(z)$ for $\abs{z}<1$ has no brunch cut, and we get a purely exponential decay.
Otherwise, for $\alpha>1$ and $\beta>1$, we get $(({\cal H}^{-1})_Q)_{2j,1}\sim \int_0^1 dx x^{j-1} (-\ln x)^{\min\{\alpha,\beta\}-1}\sim j^{-\min\{\alpha,\beta\}}$,  and for $\alpha<1$ or $\beta<1$, we get  $(({\cal H}^{-1})_Q)_{2j,1}\sim j^{-2+\min\{\alpha,\beta\}}$. We note that for $\alpha>1$ and $\beta>1$ the estimation of the decay is in agreement with Ref.~\cite{jager20}.
{\change
\section{Mutual Information: a heuristic calculation}}

To explain {\change the relation between correlation functions and long-distance mutual information}, we consider a closed chain of $L$ sites and two blocks $A$ and $B$ of sizes $L_A=L_B=\ell=x L$, separated by a distance $\ell_{AB} = y L$, with $x$ and $y$ some fraction of $L$.
{\change To investigate the correlations between the two blocks, we can} calculate the reduced density matrix $\rho_{A\cup B}$ of the two blocks. It is useful to introduce the Majorana operators, {\change as defined before}, $c_{2j-1}=a_j + a^\dagger_j$ and $c_{2j}=i a^\dagger_j - i a_j$, and since their products form a basis in the operator linear space, we can make the ansatz
\begin{equation}
\rho_{A\cup B} = \rho_A \otimes \rho_B + \frac{1}{2^{2\ell}} \sum_l (-1)^{\frac{n_l}{2}} \varepsilon_l O_l\,,
\end{equation}
where the sum is over all the possible products $O_l$ of an even number $n_l$ of Majorana operators belonging to both blocks $A$ and $B$, e.g.,
$\rho_{A\cup B} = \rho_A \otimes \rho_B - \frac{1}{2^{2\ell}} \big(\sum_{m\in A,n\in B} \varepsilon_{mn} c_m c_n
- \frac{1}{2}\sum_{m\neq m'\in A,n\neq n'\in B} \varepsilon_{mm'nn'} c_m c_{m'} c_n c_{n'} +\cdots\big)\,$
and the quantities $\varepsilon_l$ can be achieved by calculating the expectation values of $O_l$, so that we get, for instance,
$\varepsilon_{mn} = \langle c_m c_n\rangle$,
$\varepsilon_{mm'nn'} = \langle c_m c_{m'} c_n c_{n'} \rangle - \langle c_m c_{m'}\rangle \langle c_n c_{n'} \rangle$,
and so on.
If the correlation function $F_{R0}$ dominates over $C_{R0}$ for large $R$, as $L\to \infty$, the term proportional to $\varepsilon_{mn}$, which goes as $\varepsilon_{mn}\sim L^{-b}$, 
{\change is a leading term}, e.g., $\varepsilon_{mm'nn'}\sim L^{-2b}$. 
{\change For simplicity, we will consider the trivial case $\rho_A\otimes \rho_B \propto \mathbb{I}_{AB}$.}
It is easy to show that, {\change by applying the perturbation theory, up to the first order correction we get}
$\rho_{A\cup B} \sim \rho_A\otimes \rho_B - \frac{4}{2^{2\ell}} \sum_{i\in A, j\in B} F_{ij}(a_i a_j + a^\dagger_j a^\dagger_i)$.
To study the scaling with the size $L$, it is enough to consider
\begin{equation}\label{eq. scal rho}
\rho_{A\cup B} \sim \rho_A\otimes \rho_B - \frac{\varepsilon}{2^{2\ell}} \sum_{i\in A, j\in B} (a_i a_j + a^\dagger_j a^\dagger_i)\,,
\end{equation}
where $\varepsilon\sim \ell_{AB}^{-b}\sim L^{-b}$. 
By performing the transformation $a_i\to a_i^\dagger$ for $i\in A$ {\change and by summing all the perturbative corrections we get something like} $\rho_{A\cup B}\sim \exp(-\varepsilon \sum_{i,j} a^\dagger_i J_{ij}a_j)$, with $J_{ij}=1$ if $i\in A$ and $j\in B$, or $j\in A$ and $i\in B$. {\change In general, we can} characterize the correlations between the two parties $A$ and $B$, {\change  by considering} the mutual information defined as
in Eq.~(\ref{eq. mu in}). 
{\change  For the trivial case which we are considering,}
the entropy $S_{A\cup B}$ can be expressed in terms of the eigenvalues $x_j$ of the matrix $J$ as $S_{A\cup B} = \sum_j \varepsilon x_j/(1+e^{\varepsilon x_j})+\ln(1+e^{-\varepsilon x_j})$. Since there are only two non-zero eigenvalues $x_j$, which are $\pm (2xL)$, for $b>1$, $\varepsilon L \to 0$ and we get $S_{AB} \sim (2xL) \ln 2 - \varepsilon^2 (xL)^2$. Concerning the mutual information, we have $I_{A:B}\sim L^2 \varepsilon^2 \sim L^{2(1-b)}$, which tends to zero as $L\to \infty$ since $b> 1$.  However, if $b\leq 1$, $I_{A:B}$ does not vanish as $L\to \infty$, so that we have long-range entanglement in this case. In our case, looking at Table~\ref{table: decay}, we have at least $b=1$ therefore we expect that, in those cases which correspond to $\beta\le \min\{1,\alpha\}$,  $I_{A:B}$ saturates to a constant value in the limit of large system size, $L\to \infty$.
We note that this result is quite general, i.e., does not depend on the choice of $\rho_A \otimes \rho_B$.
To prove it, let us consider a general $\rho_{AB}$, not necessarily of the form of Eq.~\eqref{eq. scal rho}, and the eigenvalue equations $\rho_A \ket{\lambda^A_i} = \lambda^A_i \ket{\lambda^A_i}$, a similar equation for $B$, 
and $\rho_{A\cup B} \ket{\lambda^{AB}_{ij}} = \lambda^{AB}_{ij}\ket{\lambda^{AB}_{ij}}$, with $\lambda^{AB}_{ij}=\lambda^A_i \lambda^B_j + \delta \lambda_{ij}$. {\change  If $\delta \lambda_{ij}$ is small enough, from Eq.~\eqref{eq. mu in} it is easy to show Eq.~(\ref{eq. mi}), here reported for convenience}
\begin{equation}
I_{A:B} \sim \sum_{ij} \delta \lambda_{ij} \ln(\lambda^A_i\lambda^B_j) + \sum_{ij} \frac{(\delta \lambda_{ij})^2}{2 \lambda^A_i\lambda^B_j}\,.
\end{equation}
{\change  We already know how the eigenvalues $\lambda^A_i$ and $\lambda^B_j$ scale with $L$.} To guess the scaling of $\delta \lambda_{ij}$ {\change  with $L$}, let us consider for a moment {\change  the trivial case $\rho_{A\cup B}\sim \exp(-\varepsilon \sum_{i,j} a^\dagger_i J_{ij}a_j)$}. For the case $b\geq 1$, we have $\lambda^A_i\sim 2^{-L_A}$ and $\lambda^B_j\sim 2^{-L_B}$, and since $\sum_{ij} \delta\lambda_{ij}=0$, we get {\change $\delta \lambda_{ij} \sim \lambda_i^A \lambda_j^B\varepsilon L$}. Similarly, for $b\leq 1$, we get {\change $\delta \lambda_{ij} \sim \lambda_i^A \lambda_j^B$}. In general, we assume that the relation {\change $\delta \lambda_{ij} \sim \lambda_i^A \lambda_j^B (\varepsilon L)^\delta$} holds also for generic $\rho_{A}\otimes \rho_B$. 
In particular, for $b\leq 1$ one can expect $\delta=0$ and for $b>1$ we have verified this hypothesis numerically in our model for the largest eigenvalue $\lambda^{AB}_1=\lambda^{AB}_{11}$ and $\alpha>1$ and $\beta>1$ finding $\delta\approx 2$ (see Fig.~\ref{fig:hyp}).
Then, if the hypothesis on the scaling of $\delta \lambda_{ij}$ is true, {\change from Eq.~\eqref{eq. mi}} it is easy to see that the scaling of the mutual information is of the form
\begin{equation}\label{eq. mi scal}
I_{A:B} \sim c_1  L^{\delta(1-b)}+ c_2 L^{2\delta(1-b)}\,,
\end{equation}
where we have considered an algebraic decay with same exponent $\gamma$ for the eigenvalues $\lambda^A_i$ and $\lambda^B_j$, e.g.,  $\lambda^A_i\sim a_i L^{-\gamma}$, in the presence of long-range entanglement. 

\bigskip

{\change 
\section{Correlation matrices}
Let us consider the general case where ${\cal C}_{AB}$ is not necessarily equal to ${\cal C}_{BA}$. In this case we need to diagonalize a block matrix of the form
\begin{equation}
M=\left(
    \begin{array}{cc}
      A & X \\
      Y & A \\
    \end{array}
  \right)
\end{equation}
where $A$, $X$ and $Y$ are square matrices. To calculate the eigenvalues of $M$, we consider $X$ and $Y$ as small perturbations, thus the matrix $M$ can be expressed as $M=M_0+M_1$, where
\begin{equation}
M_0=\left(
    \begin{array}{cc}
      A & \frac{X+Y}{2} \\
      \frac{X+Y}{2} & A \\
    \end{array}
  \right)\,, \quad M_1=\left(
    \begin{array}{cc}
     0  & \frac{X-Y}{2} \\
      -\frac{X-Y}{2} & 0 \\
    \end{array}\,.
  \right)
\end{equation}
By considering the eigenvalue equation $A\ket{a_n}=a_n \ket{a_n}$, the eigenvalues $m^{(0)}_{n,\pm}$ of $M_0$ are the eigenvalues of the matrices $A \pm (X+Y)/2$, thus $m^{(0)}_{n,\pm} \approx a_n \pm \bra{a_n}(X+Y)\ket{a_n}/2$. The eigenvalues $m_{n,\pm}$ of $M$ are $m_{n,\pm}=m^{(0)}_{n,\pm}+ \delta m_{n,\pm}$, where by using the non-degenerate quantum perturbation theory, we get
\begin{equation}
\delta m_{n,\pm} \approx \frac{1}{4}\sum_{k\neq n} \frac{|\bra{a_k}(X-Y)\ket{a_n}|^2}{a_n-a_k}\,.
\end{equation}
Then, at the first order $m_{n,\pm} \approx a_n \pm \bra{a_n}(X+Y)\ket{a_n}/2$, so that in general $\delta n_q \approx (U^\dagger ({\cal C}_{AB}+{\cal C}_{BA})U)_{qq}/2$.\\
We note that if $\eta=1$, this first order correction is in agreement with the scaling hypothesis $\delta\lambda_{ij} \sim \lambda^A_i \lambda^B_j L^{2(1-b)}$, for instance let us consider the case $\delta \lambda_{11}$. We have $\lambda^{A\cup B}_1 = 1/Z_{A\cup B}$, and
\begin{eqnarray}
\nonumber
\lambda_1^{A\cup B} &=& \prod_k (n^A_{k-}-\delta n_{k-})(n^A_{k-}+\delta n_{k-})=\prod_k ((n^A_{k-})^2-(\delta n_{k-})^2)\\
\nonumber&\approx& \prod_k (n^A_{k-})^2 - \sum_k (\prod_{k'\neq k} (n^A_{k'-})^2)(\delta n_{k-})^2\\
&\sim& (\lambda^A_1)^2 + (\lambda^A_1)^2 l^{2-2b}
\end{eqnarray}
from which $\delta \lambda_{11}\sim \lambda_1^A \lambda_1^B l^{2-2b}$, since $\lambda^B_i=\lambda^A_i$.
}

{\change 
\section{Kibble-Zurek mechanism}}

{\change Let us consider} a chemical potential which changes linearly in time as $\mu(t)=t/\tau_Q$ for $t\in(-\infty,0)$, such that we cross the quantum critical point $\mu_c=-w g(0)$ for $\beta>1$ {\change and $\alpha>1$.} At the initial time $t=-\infty$, the initial state is the state $\ket{\psi(-\infty)}$ defined such that $a^\dagger_k \ket{\psi(-\infty)}=0$ for any $k$. At the final time $t=0$, we get $\mu(0)=0$ and the Hamiltonian can be expressed as $H=\sum_k  \epsilon_k \alpha_k^\dagger \alpha_k$. The state at the final time is $\ket{\psi(0)}$, and we focus on the excitations probability $p_k = \bra{\psi(0)}\alpha^\dagger_k\alpha_k\ket{\psi(0)}$, and the excitations density $n = \int_0^\pi dk p_k/\pi$. This can be calculated by using the time-dependent Bogoliubov method (e.g., see Ref.~\cite{Dziarmaga05}), from which we get Landau-Zener differential equations. For $\beta>1$ the gap closes at $k=0$ and only large wavelengths contribute for which we get $p_k \sim \exp(-\pi \tau_Q (\Delta f(k))^2)$, so that, if $1<\beta<2$, $f(k)\sim k^{\beta-1}$ as $k\to 0$ and we get $n\sim \tau_Q^{1/(2-2\beta)}$, and if $\beta>2$, $f(k)\sim k$ and $n \sim 1/\sqrt{\tau_Q}$, in agreement with Ref.~\cite{defenu19}.
By considering the Kibble-Zurek mechanism~\cite{zurek85}, if the so-called healing length goes like $\xi \sim|\mu-\mu_c|^{-\nu}$ and the gap closes as $E_{gap}\sim |\mu-\mu_c|^{z \nu}$, so that the relaxation time is $\tau_{rel}\sim 1/E_{gap}$, we expect the density of defects $n \sim \tau_Q^{-\nu/(1+z \nu)}$. In our case, the gap closes with $z\nu =1$, thus we expect $\nu=1$ for $\beta>2$ and $\nu=1/(\beta-1)$ for $1<\beta<2$.
Conversely, by considering $\mu(t)=-t/\tau_Q$ for $t\in(-\infty,0)$, so that we cross the quantum phase transition point $\mu_c=-w g(\pi)$, we have a contribution only for $k$ near $\pi$, otherwise the tunnelling is negligible. Since $f(k)\sim (k-\pi)$ as $k\to \pi$, we get $n \sim 1/\sqrt{\tau_Q}$ for any $\alpha$ and $\beta$, and from the Kibble-Zurek theory we expect $\nu=1$.

\end{document}